\documentstyle[12pt,aasms4]{article}

\received{}
\accepted{}

\slugcomment{submitted to AJ}

\begin{document}

\title{Near-infrared Observations of the Proto-Planetary Nebula \\
IRAS 07131-0147}

\author{David R. Alves\altaffilmark{1} }
\affil{Institute of Geophysics and Planetary Physics, 
Lawrence Livermore National Laboratory, Livermore, CA 94550 \\
Electronic mail: alves@igpp.llnl.gov }
\altaffiltext{1}{Department of Physics, University of California, 
Davis, 95616}

\author {D. W. Hoard}
\affil{Department of Astronomy, University of Washington, 
Seattle, WA 98195 \\
Electronic mail: hoard@astro.washington.edu }

\begin{abstract}

We have obtained near-infrared imaging and an optical spectrum
of the proto-planetary nebula IRAS~07131$-$0147, 
a highly polarized bipolar reflection nebula 
believed to be in evolutionary transition from the 
asymptotic giant branch to the planetary nebula phase.  
Our images reveal
point reflection symmetry in the lobes - a relatively rare 
morphological feature.
We place an upper limit on the
distance of 6.5 kpc. 
Utilizing numerical integrations of single grain 
scattering models we find the nebula to lie at an inclination angle of 
$i=20\pm5$ degrees in the plane of the sky.
We present a refined geometric interpretation of IRAS~07131$-$0147 
consistent with our new observational data and argue that the
central star is likely to have an as yet undetected binary companion.

\end{abstract}

\keywords{proto-planetary nebulae: individual (IRAS~07131$-$0147) $-$ stars: 
binaries, late type, mass loss, evolution $-$ infrared: sources }

\section{Introduction}

Proto-planetary nebulae (PPN) occur in the poorly understood transition
between the asymptotic giant branch (AGB) and planetary nebula (PN) 
stages of stellar evolution.  As a star moves up the AGB, it loses mass 
at an increasing rate, culminating in a final cataclysmic superwind
(\cite{Wood 93}).  The end of mass-loss signals the transition to the 
beginning of the PPN phase 
(\cite{Kwok 93}).  At this point, the central star with a typical core mass 
of $M_{c} = 0.6M_{\odot}$ and temperature of $T_{*} \approx 5000$ K 
(\cite{Volk and Kwok 89})
evolves blueward in the HR diagram, its temperature increasing as it loses 
additional envelope mass through hydrogen shell burning, until it reaches 
$T_{*} \approx 30,000$ K necessary to ionize the circumstellar shell and 
produce a PN.  
The ionization 
of the nebula defines the end of the PPN stage.  
The typical ages of PN, 
the large dust densities required for their formation, and the minimum 
central star temperature needed to ionize the ejected gas strongly constrain 
the duration of the PPN phase to a few thousand years (\cite{Renzini 83}). 

Some PPN have easily discernable bipolar morphologies reminiscent of
``butterfly'' PN, in which the central star illuminates two diametrically
opposed, symmetric lobes of nebulosity (\cite{Balick 87}).
Butterfly PN structure is generally attributed to a density
contrast characterized by the presence of a higher concentration
of dust in an
equatorial plane around the central star and an interacting wind,
where a hot wind plows into a previously ejected, slow moving
envelope (\cite{Livio 93}).
The morphological similarity between
bipolar PPN and butterfly PN, and their
evolutionary relation, allow us to interpret the objects
in an analagous manner.  Morris (1981) investigated the
dust distributions that would reproduce bipolar reflection nebulae and
found that a density contrast was a necessary criterion.
A waist of dust blocks the light from the star
in the equatorial region, while
starlight emerging into the lower density lobes of the polar regions
experiences less extinction.
His dust distributions reproduce the ``horns'' seen in some bipolar
PPN; they are caused by changes in the optical depth structure of the lobes.
The equatorial density enhancement appears well founded.  Observational
evidence indicates that such dust distributions begin forming well
before the PPN stage (\cite{Johnson and Jones 91}).  However,
full development of lobe structure, in agreement with the
scattering models of Morris, and by analogy with the butterfly
PN scenario of Balick, requires an interacting wind.  Herein lies
one of the most puzzling aspects of PPN; namely, where does the second wind
come from?  If we accept, for the moment, that some mechanism is capable
of producing a fast wind and we account
for the rapid evolution of the post-AGB central stars, then PPN may lie
along a morphological sequence related to their age/evolutionary state.
They evolve from
young systems with totally obscured central stars surrounded by reflection
nebulae (e.g. AFGL 2688, OH 0739$-$14),
to middle-aged systems in which the circumstellar material has been
sufficiently dispersed to reveal the central star (e.g. Frosty Leo),
to old systems in which
the central star has begun to photoionize the surrounding nebula,
which is then
seen in both reflection and emission (e.g. M2-9, Mz-3)
(for examples see \cite{Bujar 92} and references therein).
This sequence is admittedly
based on a small sample of transition objects whose ages are poorly known.
Furthermore, some PPN have completely obscured central stars and
photoionized central cavities (e.g. AFGL 618; \cite{Kwok 93}); possibly 
evolving so quickly as to appear with both young and old features.
Nevertheless, we proceed
with our data interpretation in the context of this intuitive, albeit
loosely defined, evolutionary sequence.

Wolstencroft, Scarrott, and Menzies (1989) discovered a bipolar
reflection nebula at the position of the infrared object IRAS~07131$-$0147
(galactic coordinates l=217$^{\circ}$, b=+4.7$^{\circ}$) during a search 
for optical counterparts of IRAS sources
with peak flux densities at $25~\micron$.  They classified the nebula as a
PPN of intermediate age based on the fact that the lobes are seen in
reflection only and the central star is relatively unobscured.  The
central star was found to be oxygen-rich and classified as spectral
type M5 III based upon strong TiO bands in its optical spectrum and a
$10~\micron$ silicate feature in the IRAS $Low$ $Resolution$ $Spectrum$
(LRS) (Scarrot et al.~1990, hereafter S90).  
The IRAS LRS is consistent with PPN model
spectra; furthermore, the rising flux from 10 to
20 $\micron$ indicates that IRAS~07131$-$0147 is somewhat evolved
(\cite{Volk and Kwok 89}).  The latter feature
is characteristic of an evacuated cavity created as the
dust shell expands away from the central star and cools.
The optical morphology and IRAS data suggest that the nebula has
evolved considerably, yet the central star has not started moving
toward hotter temperatures as single star evolutionary calculations
predict.  In this paper, we present near-infrared imaging and an
optical spectrum of IRAS~07131$-$0147 that allow us to refine our
understanding of the physical nature of this interesting transitional
object.

\section{Near-infrared Observations}

On 1995 April 16 UT 
we obtained near-infrared images of IRAS~07131$-$0147 under photometric
conditions using the Lick Observatory 3m telescope and the LIRC II
mercury-cadmium-telluride 256$\times$256 pixel camera.  Broadband $J$
(1.10 to 1.44 \micron) and $K^\prime$ (1.95 to 2.35 $\micron$) filters
were utilized.  The intermediate field-of-view optical configuration on
the telescope yielded a pixel size of $0.36~\arcsec$ and full image
area covering 1.62$^\prime$$\times$1.62$^\prime$ which provided
good sampling of the psf in the 1.5 arcsec seeing conditions.
We implemented an observing strategy of 30 and 15 second exposures 
in a four-point,
on-source dithering pattern ($J$ and $K^\prime$, respectively), optimally
exposing the target star and providing the images needed to construct
sky flats.  We derived coordinate offsets for each image from the
centroids of several bright stars.  The sky flattened images were then
offset and median combined to create final, 150 second cumulative
exposure images in both $J$ and $K$.  The $J$ image is presented in 
Fig.~1 (plate 1).
Transformation equations from
instrumental to standard $J$ and $K$ 
magnitudes, including a zero point offset and
extinction coefficients, were calculated
from concurrent observations of
three standard stars in the UKIRT faint standard list taken
at airmasses ranging from 1.0 to 2.1.  
We find the central star in IRAS~07131$-$0147 to have $J$ = 9.92 $\pm$ 0.10
mag and $K =9.05\pm 0.10$ mag.
We compared the stellar profile of the PPN central star to field stars in our 
images and found no evidence for a non-stellar
profile in the former.  In magnitudes per pixel, the lobes have peak
intensities of 
$J_{west} = 19.95$, $K_{west} = 19.85$, 
$J_{east} = 19.05$, and $K_{east} = 18.60$, with uncertainties
of order 0.1 to 0.2 mag.

\subsection{Morphology}

The eastern lobe is
brighter and more extended than the western lobe.  This was also
observed by S90 in optical images and suggests a non-zero angle 
of inclination relative to the plane of the sky.  On the
other hand, the similar level of polarization found in both lobes by
S90 suggests an inclination near zero.  
We return to this in \S 4;  however, 
we note that if
the difference in brightness between the
eastern and western lobes is due to
the inclination of the nebula, then one expects
the fainter lobe 
to be redder than the brighter lobe.  Synthetic aperture photometry
(beam diameter = 18$\arcsec$; central star subtracted)
performed on the east and west lobes in the $J$ and $K$ images of 
IRAS~07131$-$0147 
indicates that the faint western lobe 
($J-K = 0.1\pm0.2$) is bluer than the bright eastern lobe 
($J-K = 0.4\pm0.2$).  
IRAS~07131$-$0147, like 
IRAS~09371+1212 (\cite{Roddier et al. 95}) and OH~0739$-$14 
(\cite{Cohen et al. 85}),
is a PPN whose lobe colors
demand a more sophisticated geometrical interpretation.

In contrast to the optical images of S90, which show bright ridges of clumpy
knots located approximately 7 arcseconds to the east and west of the
central star, our $J$ image shows peak
intensities inside this region.  The nebulosity in the $K$ image
(not shown) does not extend quite as far from the central star
as in the $J$ image, but is otherwise
very similar.  Using the midpoint of the horns at the
outer edges of the nebula to define the polar axis, we find deviations
from plane symmetry inside the ridge region.  The peak intensity levels in
the eastern lobe bend slightly to the north while
in the western lobe they
bend to the south.  In Fig.~2 we have plotted the $J$ band intensity profiles
perpendicular to the polar axis at 2 pixel intervals to the east and
west of the central star.  
These profiles
clearly demonstrate 
evidence for point reflection
symmetry near the central star.  
This is a relatively rare characteristic in bipolar nebulae 
and can possibly be understood in the context of a precessing jet in
a binary system (\cite{Morris and Reipurth 90}).

\subsection {Distance}

We find no distance for IRAS~07131$-$0147 in the
literature and have attempted to estimate one from
our $K$ photometry of the central star.  
A sample of 14 M giants in Baade's window were observed
by Frogel and Whitford (1987).  They found a median $K$
magnitude of 9.2 with a spread of $\delta K \approx \pm 0.7$ mag.
Using their distance modulus of 14.2, we adopt 
the absolute $K$ magnitude
calibration of $-5.0~\pm$ 0.7 mag for an M5 giant.  We ignore the
possible effect of a high bulge metallicity in this analysis and note
that the advantage of using a reasonably large sample of M5 giants at
a known distance instead 
of a single calibrating object outweighs this concern.
This yields a distance of 6.5 $\pm$ 3 kpc for IRAS~07131$-$0147.  
The ($J-K$)
color of +0.87 for the central star is slightly
bluer than the median (de-reddened) color of the sample of
galactic M giants ($J-K$ = +0.97), implying negligible
reddening.  This is also consistent with 
the nebula being in a generally
evolved state.
The neutral hydrogen
density at this galactic latitude and longitude would
contribute negligible foreground reddening in the $K$ band.  
We note that applying our
magnitude calibration to the published infrared photometry
of OH~0739$-$14, which also contains an M giant
central star, yields a distance for that
PPN in agreement with previous estimates (\cite{Woodward et al. 89}, 
and references therein).  
Clearly, however, these are 
not ``normal'' M giants.
Given the peculiar nature
of the central star and its poorly understood evolutionary state
(\cite{Iben and Tutukov 93}),
we consider
our distance estimate for IRAS~07131$-$0147 an upper limit. 

At a distance of 6 kpc, the angular extent of the east
lobe in the optical images of S90 corresponds to an approximate size
of 1.2 pc.   Molecular line profiles indicate bipolar wind velocities
of order tens to hundreds of km~sec$^{-1}$ in PPN (\cite{Bujar 92}).  
Adopting a wind
velocity of 100 km~sec$^{-1}$, we estimate a dynamical age 
for IRAS~07131$-$0147 of around 
12,000 years.  
Under the same assumptions, the east lobe as seen
in our near-infrared image would have
a dynamical age of approximately 5000 years.  
Although the uncertainty in the
age is probably
a factor of two or three, it is consistent with the general properties
of this PPN; in particular, the evolved dust distribution.
Adopting the standard ruler assumption for PPN ($7\times10^{15}$ m;
\cite{BB 91})
gives a distance of 1.3 kpc and an age of 2300 years.  The central star, in
this case, would be 3.6 magnitudes fainter than the median absolute $K$ 
magnitude of galactic bulge M giants.

\section{Optical Spectrum}

We obtained a single, 10 min exposure, low resolution (8~\AA\ in blue,
12~\AA\ in red) optical spectrum of IRAS~07131$-$0147 on 1995 December 6
UT using the Double Imaging Spectrograph (DIS) 
on the Apache Point Observatory (APO) 3.5m
telescope.  After passing through a $6'\times1''$ slit,
incoming starlight is split into two separately-collimated beams by a
dichroic with a transition wavelength of 5350~\AA.  Each beam then
passes through a separate grating and falls onto a detector.  The
``blue'' side utilizes a 150 line/mm grating with a thinned, UV-coated
$512\times512$ SITe CCD; the ``red'' side utilizes a 300 line/mm
grating with a thinned $800\times800$ TI CCD.  The grating tilts were
set to give wavelength ranges of 3800--5800\AA\ on the blue side and
5800--10000~\AA\ on the red side.  The spectrograph slit was aligned 
with the long axis of IRAS~07131$-$0147 for the exposure.  
The spectrum of the central star, as shown in Fig.~3,
is that of a late-type giant
(as reported by \cite{WSM 89} and S90), 
showing prominent TiO absorption bands.  
The spectrum of the eastern lobe 
is almost identical, attesting to the fact that most of the 
light emerging from the lobe is scattered from the central star.  
The fainter western lobe was not detected.

\section{Inclination}

Numerical models in which each emergent photon is singly scattered by 
a dust grain in the nebula have been used to reproduce the observed 
intensity contours of PPN and to estimate their inclinations 
(e.g., \cite{Morris 81}, \cite{Latter et al. 92}, \cite{Latter et al. 93}).  
IRAS~07131$-$0147 displays a very high degree of polarization 
(60--70\%; \cite{WSM 89}), suggesting that the 
number of photon scatterings is small (\cite{Latter et al. 93}) 
and recommending the single-grain scattering method.  
We followed the procedure of Morris (1981), 
integrating 
over an emission coefficient per unit volume due to scattering
along the line-of-sight 
to the nebula, projected onto a grid in the plane of the sky.  
The scattering is assumed to be isotropic and we include absorption
of the scattered light along the line-of-sight.  The dust
distribution is that produced by a
constant velocity outflow of dust in the nebula, 
modified so
that the density falls off 
rapidly beyond some critical latitude (ensuring that the 
nebula becomes optically 
thin at high latitudes).  This approximates a low density ``cone''
along the polar axis in each lobe shaped by a fast wind from the
central star.

We used the dust distribution of Morris's 
model~1d  (1981), which decreases exponentially in latitude with an e-folding
length of $30^{\circ}$ and a cut-off latitude of $60^{\circ}$.
This model provides a 
reasonable reproduction of the observed PPN.  The theoretical intensity
contours for the adopted dust distribution at four different inclination
angles are shown in Fig.~4.
The model is clearly idealized in several respects.  For example, it does
not reproduce the point symmetry in the inner lobe regions.  Furthermore, it
does not reproduce the enhanced ``ridges'' seen prominently in the optical
images and less so in our near-infrared images.  
After experimenting with a number of dust distributions modified 
from model~1d, 
we found that the inclination determined from an axial profile was not a 
strong function of the dust distribution among similar distributions. 
In short, a simple polar axis intensity profile best measures the
inclination while intensity contours best constrain the dust
distribution.   
Profiles with zero inclination have lobes of equal intensity.
Increasing the inclination angle increases the relative difference
in intensity between the lobes.  This is shown in Fig.~5, where
we present the polar axis intensity profiles corresponding to the different
intensity contours of Fig.~4.

From an observational point of view, we can ``smooth over''
those morphological features in our images which cannot
be reproduced by the model nebulae and measure the inclination
using the polar axis intensity profile.  The critical
assumption here is that the peak intensities of reflected light in each lobe
are primarily due to the optical depth structure and not absorption by
intervening dust (the equatorial waist).
The observed peak of the west lobe 
is 35--40\% as bright as the east lobe in our $J$ and $K$ images,
which corresponds to an inclination of $i = 20^{\circ} \pm 5$.
The observed normalized intensity profile of
IRAS~07131$-$0147, obtained by summing a 10-pixel wide strip along the
polar axis of our $J$ image, and the model profile (for $i = 20^{\circ}$),
is shown in Fig.~6.  The central star is denoted by a dashed line, emphasizing
the fact that the model does not account for the evolved, un-obscuring nature
of the equatorial dust near the central star.  The model nebulae only
require that the equatorial dust be of higher density than the lobes
(\cite{Morris 81}).

S90 find an inclination of zero from their optical polarization
study $-$ outside our most conservative error estimate.
The difference can be explained by precession of
the fast wind, where in addition to rotation
in the plane of the sky (as seen in Fig.~2), 
the fast wind has also rotated around the
equatorial axis.  Our near-infrared images probe the inner regions
of the nebula while optical data are sensitive to the outer
lobe regions (see polarization maps of S90).  In effect, we are
looking at the geometry of dust shaped by a precessing wind
at different times.
The sense of
precession is that of the fast wind in the east lobe rotating toward us
and slightly to the north.  A precession of 20 degrees in 5000 years
corresponds to an angular velocity of approximately $7\times10^{-5}$
radians/year. 
If we follow this line of reasoning farther back in time, the
east lobe may once have been the fainter lobe $-$ inclined
away from us.  Observed now, the optical depth in the extreme
outer regions of the lobes would be too small to scatter light
and the wind-shaped cone at this original inclination would not
be seen.  However, the equatorial waist would still be tilted 
over the east lobe and away from the west lobe, 
thus reddening light that is preferentially scattered
into the line-of-sight by the present day geometry 
of the inner region of the east lobe and explaining the 
observed lobe colors.  
The ridge-like brightening of the lobes several arcseconds away
from the central star would likely be a geometric effect; a result of
the curving low-density polar cones and not a physical clumpiness
in the dust.

\section{ Conclusion }

Morris (1981, 1987) proposed a binary star bipolar jet collimation mechanism 
which invokes a bound disk excreted from the red giant and a wind emanating
from accretion onto the secondary.  
This model is capable of producing fast winds 
even for PPN with late type central stars. 
Multi-dimensional hydrodynamic studies have shown that dust distributions
consistent with the formation of bipolar reflection nebulae 
are an inevitable consequence of AGB mass ejection in common envelope
binary systems (\cite{Livio 93}).  
In this scenario, the excretion of
the disk results from a brief period of friction-induced ``spiraling-in''
during the common envelope evolution.  Accretion, and possibly jets, would
follow this phase.  The suggestion by Livio (1993) that the secondary
may retain high entropy material, expand adiabatically,
and transfer material back onto the red giant 
could be responsible for increasing the mass of the hydrogen
envelope over its critical value (\cite{Kwok 93}) and postponing
its blueward evolution though the PPN phase.
The coupling of the orbital and rotational angular momenta
of the stars in the system may cause the precession of the jets
(\cite{Morris and Reipurth 90}).
Recent 3-dimensional hydrodynamic simulations of precessing jets
in point-symmetric nebulae conclude that interacting binary
systems are the most plausible jet production mechanism 
(\cite{Cliffe 95}).
The resolution of binary systems in 
several PPN through adaptive optics imaging
(\cite{Roddier et al. 95}) lends some
observational support to the binary mechanism.  
Objects such as IRAS~07131$-$0147 and IRAS~09371+1212 (Frosty Leo), 
that display evidence of evolved bipolar
nebular structure and late type (unevolved) central stars, are
likely the typical examples of binary PPN.

We believe the observational data now available for IRAS~07131$-$0147
supports the binary mechanism of PPN formation.  The optical 
and near-infrared images suggest
an evolved PPN morphology where the central star
is no longer obscured by its equatorial waist.  
The IRAS LRS
also indicates an evolved nebular morphology, where the dust has
expanded away from the central star and cooled.  Yet the central
star is a late M giant, virtually unevolved through the PPN phase.
We find striking evidence for the precession of the fast dust-shaping wind in 
this object.  Our near-infrared images reveal point symmetry in
the inner lobe regions.  Furthermore, a comparison with 
the inclination derived from optical polarization data, sensitive to
the outer lobe regions, and our infrared-derived inclination, sensitive
to the inner lobe regions, suggests that the fast wind also precessed
around the equatorial axis.  This dynamic geometric interpretation is 
consistent with the observed lobe colors and the ridge-like brightening
of the optical nebulosity several arcseconds away from the central star.  
No viable theory other than the binary mechanism has been proposed 
for the collimation or precession
of a fast wind in PPN systems with late type central stars.

The binary central star hypothesis could be tested in several
ways.  High signal to noise optical spectra of the reflection lobes
could reveal a faint blue bump, emanating from the unseen binary
companion.  This was 
suggested by
Cohen et al.~(1985) to explain a blue bump in the lobe spectrum
of the PPN, OH 0739$-$14; the binary companion was later resolved by 
adaptive optics
imaging (\cite{Roddier et al. 95}). 
The $Hubble$ $Space$ $Telescope$ may be able to resolve a companion 
in IRAS~07131$-$0147.
Alternatively, if the binary system is eclipsing, a photometric
monitoring campaign could confirm our hypothesis.

\section { Acknowledgments }

This research was partially supported by a Sigma Xi Grants in
Aid of Research Award.  Alves' research
at a DOE facility is supported by an Associated Western Universities
Laboratory Graduate Fellowship.  
Work at LLNL performed under the auspices of the USDOE contract
no. W-7405-ENG-48.
Hoard's research is supported by NSF grant no. AST 9217911.
We thank our respective advisors,
Kem Cook and Paula Szkody, and acknowledge
our respective institutions, IGPP and U. Washington, for
their support.
We thank
Bruce Balick, Robert Becker, Bill Hora, and Ed Moran for their
useful comments and discussions.

\newpage

\begin{figure}
\caption{The $J$ band image of IRAS~07131-0147 obtained on 1995 April 16 UT
using the LIRC II camera on the Lick Observatory 3m telescope.  The overlayed
contours are in 3 sigma intervals above the sky.}
\end{figure}

\begin{figure}
\caption{$J$ band intensity profiles perpendicular to the polar axis at
two pixel intervals in the East and West lobes of IRAS~07131-0147.  These
show the point reflection symmetry in the inner lobe regions.}
\end{figure}

\begin{figure}
\caption{Optical spectra of the central star and bright, eastern lobe of
IRAS~07131-0147 obtained on 1995 December 6 UT using the Double Imaging
Spectrograph on the Apache Point Observatory 3.5m telescope.  The spectrum
of the lobe has been offset downward for clarity.}
\end{figure}

\begin{figure}
\caption{Intensity contour plots at four inclinations for the adopted
single grain scattering model nebula.  The contours are
spaced at intervals of 0.5 mag.}
\end{figure}

\begin{figure}
\caption{Polar axis intensity profiles at four inclinations for the
adopted single grain scattering model nebula. They have been normalized to
a maximum intensity of one.}
\end{figure}

\begin{figure}
\caption{The observed normalized polar axis intensity profile of
IRAS~07131-0147 obtained by summing a 10-pixel wide strip in our
$J$ band image.  The central star is denoted by a dashed line.
The dotted line is the polar axis intensity profile of our
model nebula at an inclination of $i = 20^{\circ}$.}
\end{figure}

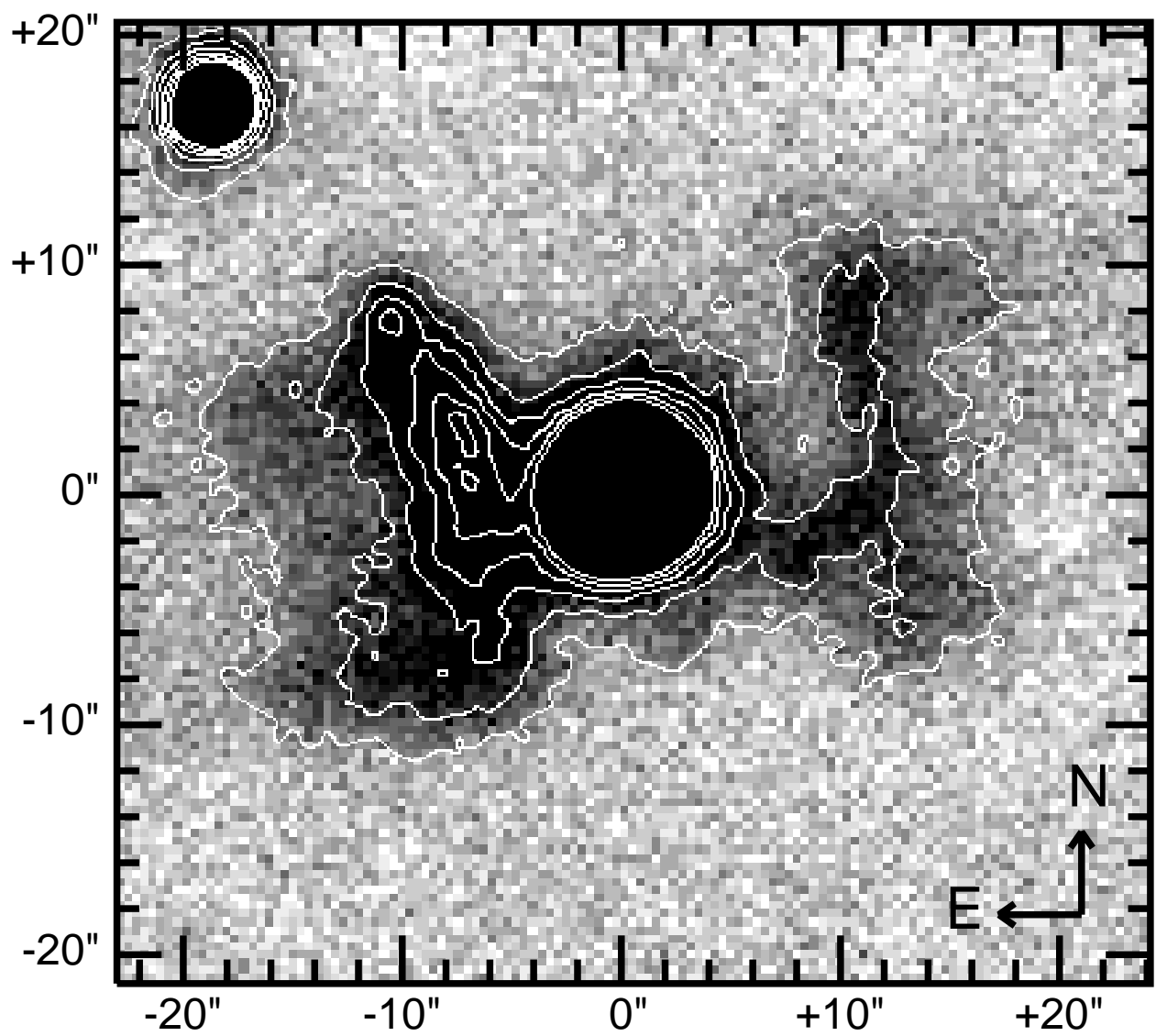
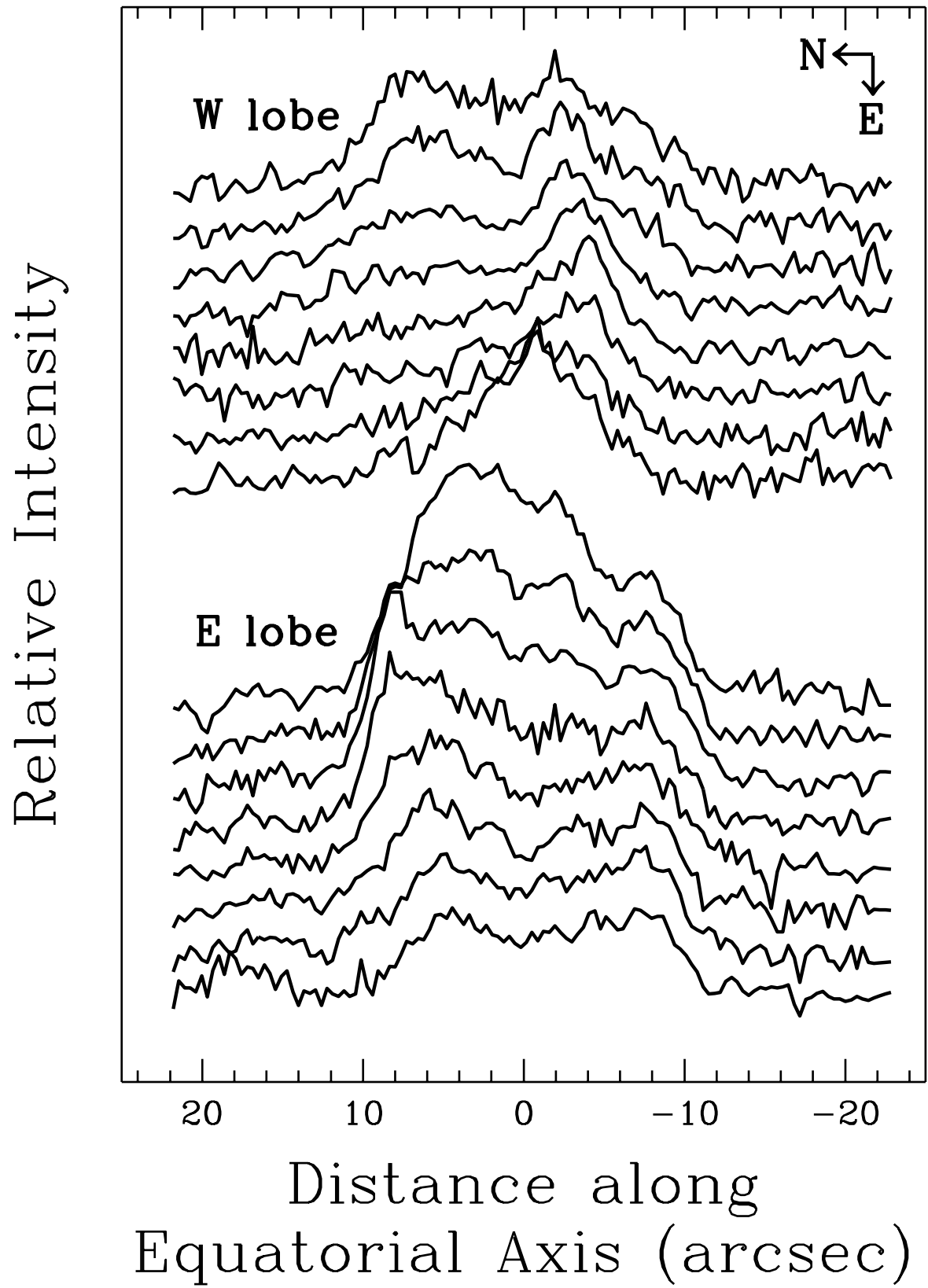
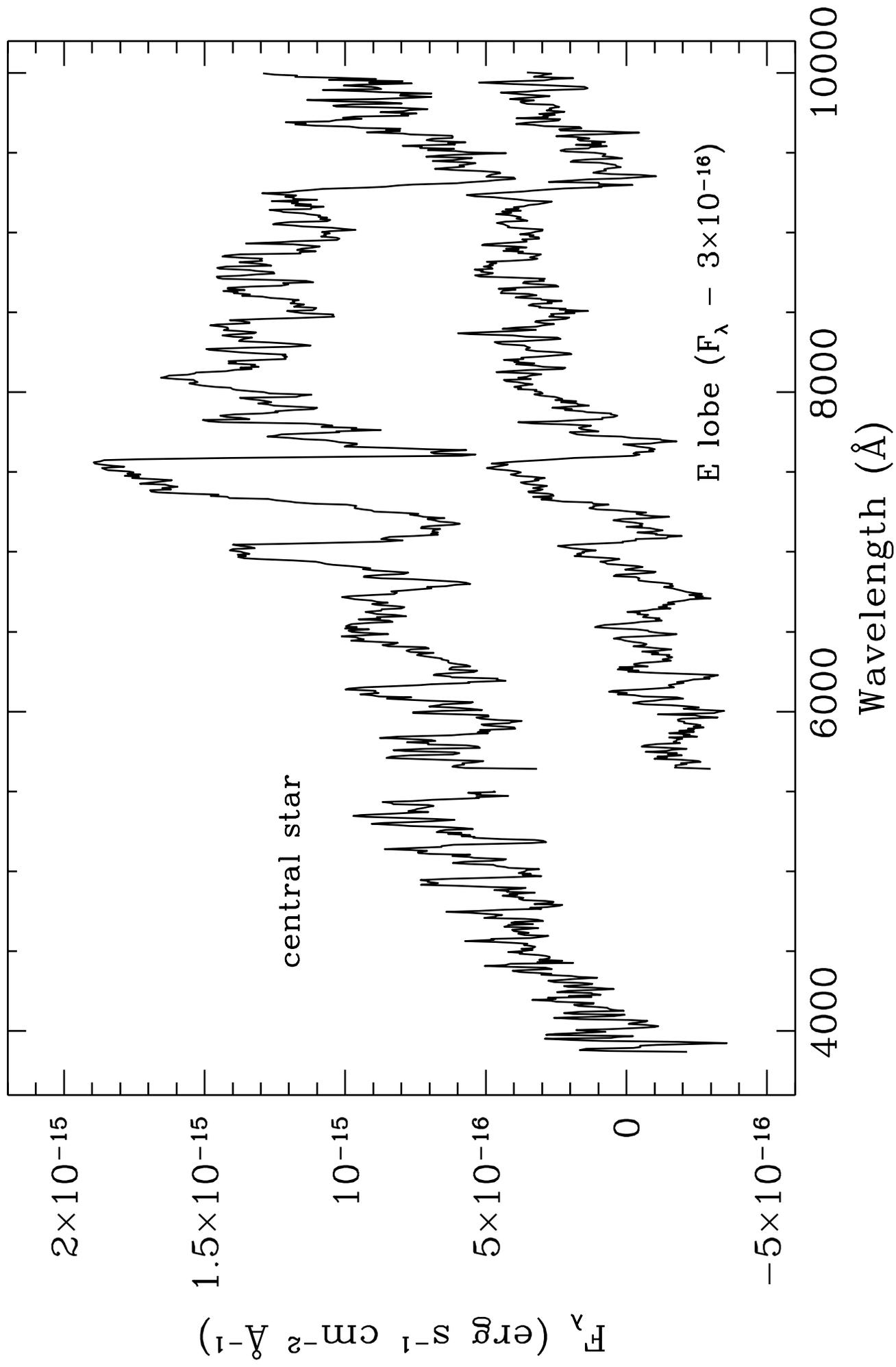
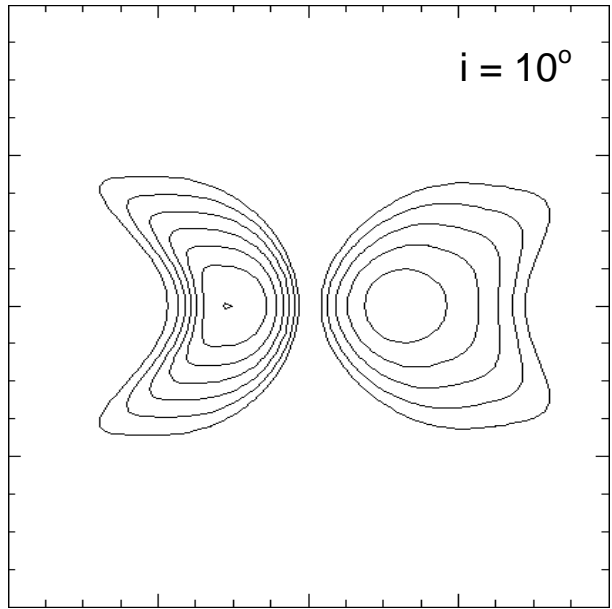
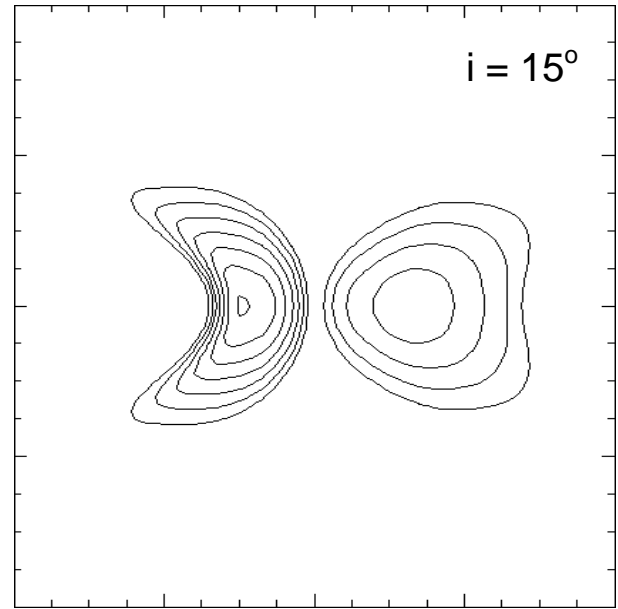
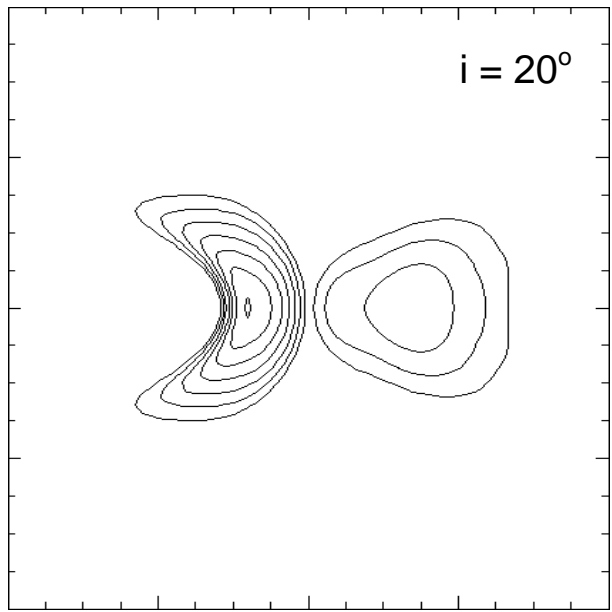
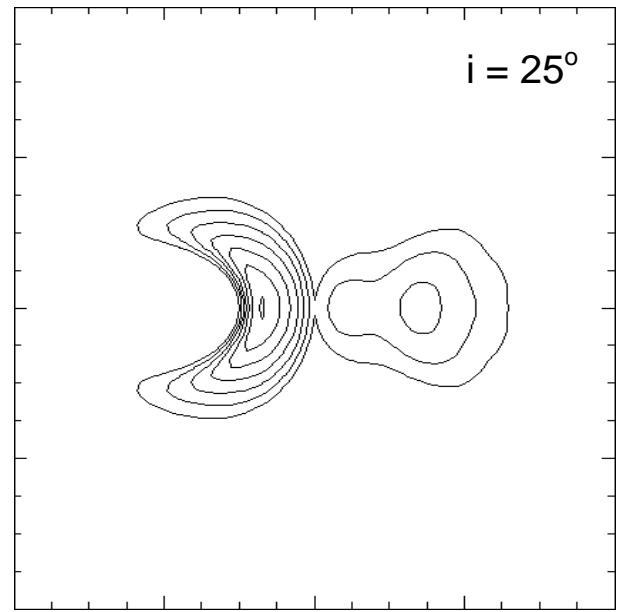
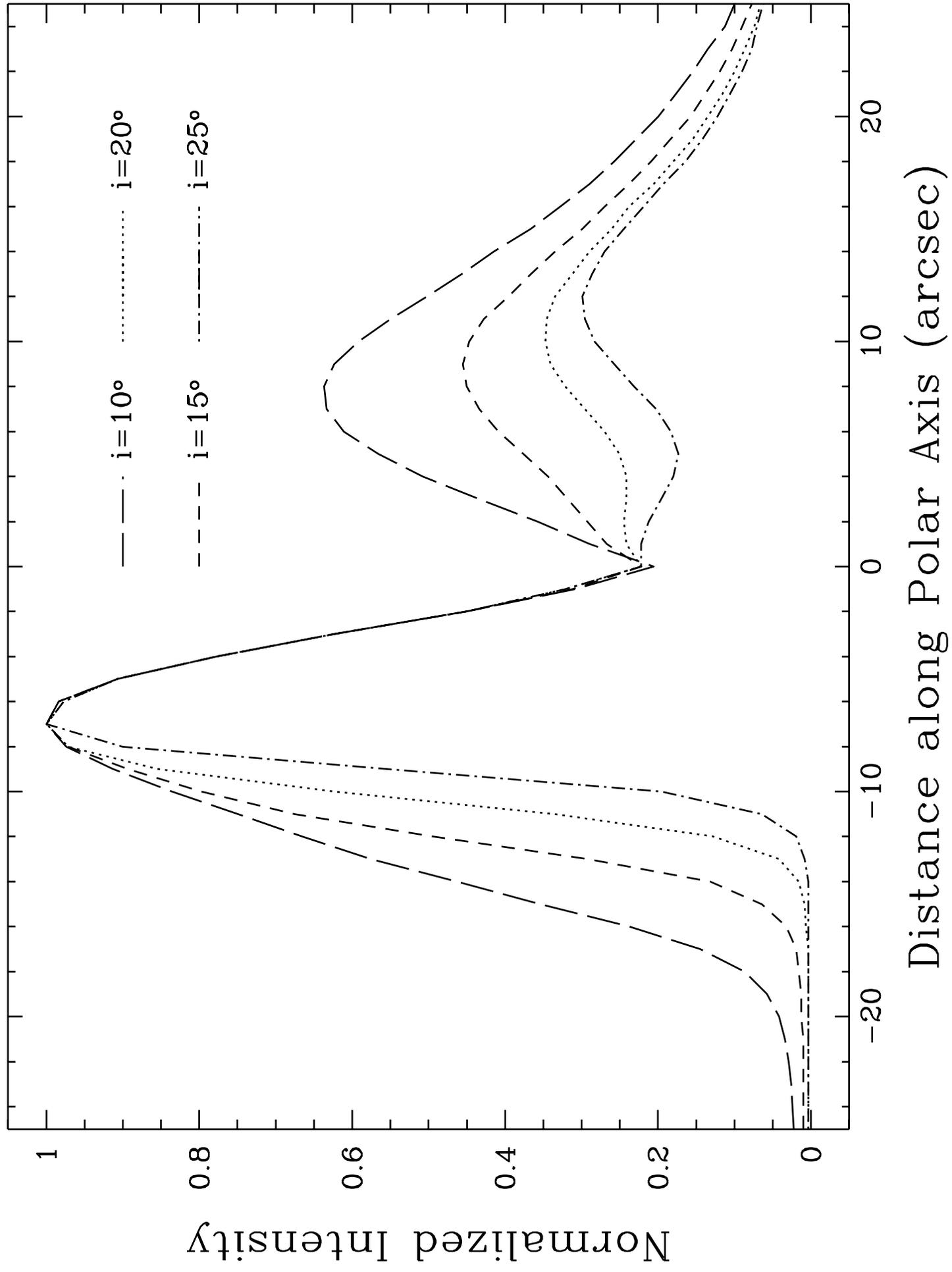
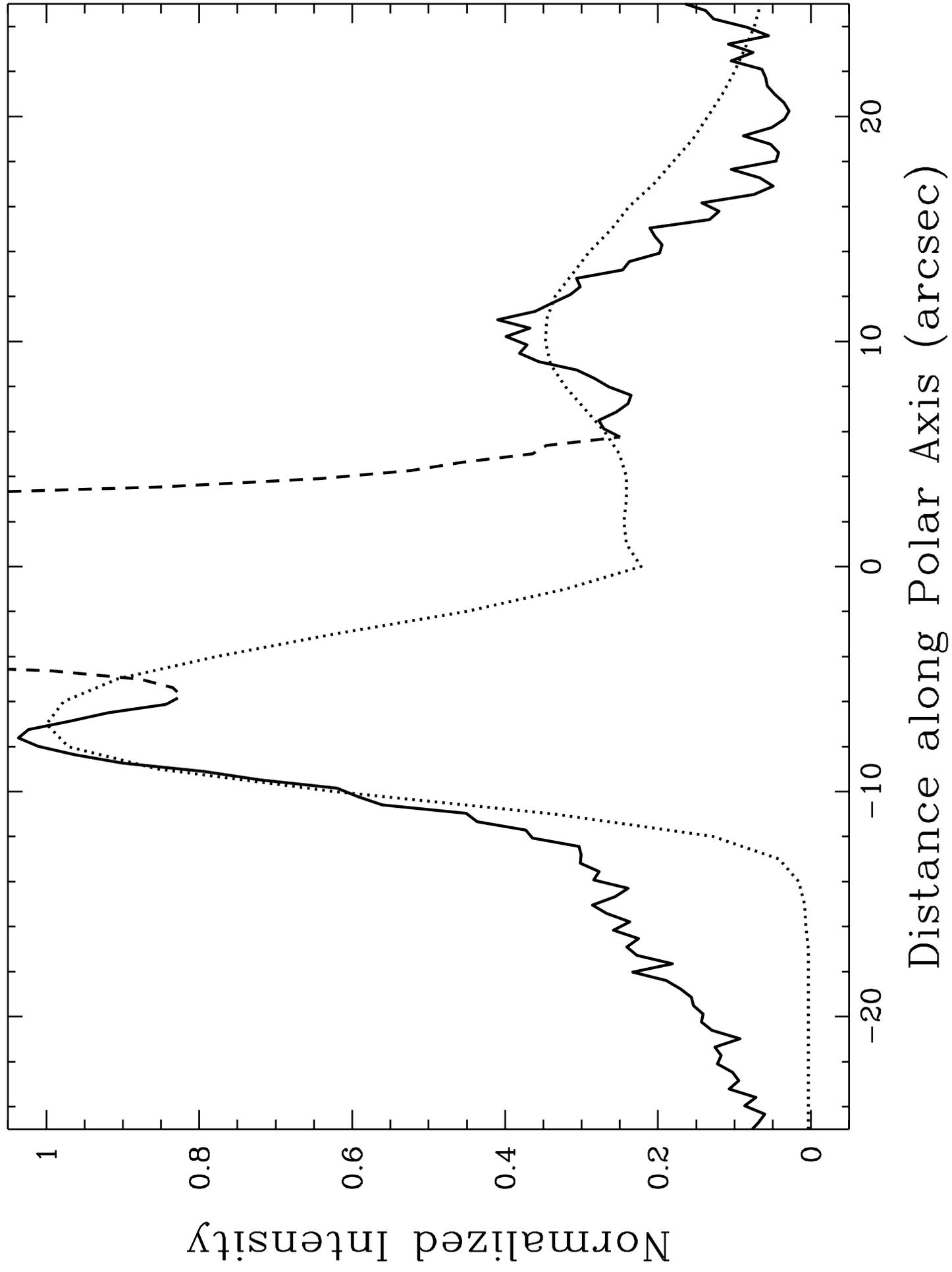
\end{document}